\documentclass[
preprintnumbers]{revtex4}
\usepackage[dvips]{graphicx}
\begin{document}
\preprint{\vtop{
{\hbox{YITP-08-43}\vskip-0pt
                 \hbox{KANAZAWA-08-05} \vskip-0pt
}}}
\title{Hidden-charm scalar tetra-quark mesons}
\author{Kunihiko Terasaki}
\affiliation{
Yukawa Institute for Theoretical Physics, Kyoto University, Kyoto 
606-8502, Japan\\
Institute for Theoretical Physics, Kanazawa University, Kanazawa 
920-1192, Japan}
\begin{abstract}
Decays of hidden-charm scalar four-quark mesons as the partners 
of $D_{s0}^+(2317)$ which has successfully been assigned to the 
iso-triplet four-quark scalar $\hat F_I^+\sim [cn][\bar s\bar n]_{I=1}$ 
meson are studied. Because OZI-rule allowed strong decays are 
kinematically limited, their radiative decays are expected to be 
important.
\end{abstract}
\maketitle

After the observation of $D_{s0}^+(2317)$~\cite{Babar-D_s} and   
$X(3872)$~\cite{Belle-X-rho}, tetra-quark mesons (including meson-meson 
molecules) are attracting general interests~\cite{Swanson}. Under this 
circumstance, we have proposed that (i) $D_{s0}^+(2317)$ is an 
iso-triplet tetra-quark scalar meson~\cite{Terasaki-D_s} and (ii) 
$X(3872)$ consists of two axial-vector meson states with (approximately) 
degenerate masses and opposite $G$-parities, which can be realized by 
tetra-quark system~\cite{Terasaki-X} without violating conservation of 
isospin and $G$-parity. As the next step, therefore, we study 
hidden-charm scalar tetra-quark mesons in this short note. 

Before visiting hidden-charm scalar mesons, we here review very briefly 
four-quark meson states. First of all, they can be classified into the 
following four groups~\cite{Jaffe}, 
\begin{equation}
\{qq\bar q\bar q\} = 
[qq][\bar q\bar q] \oplus (qq)(\bar q\bar q) 
\oplus \{[qq](\bar q\bar q)\pm (qq)[\bar q\bar q]\},
                                                 \label{eq:4-quark}
\end{equation}
according to symmetry property of their flavor wavefunction, where 
parentheses and square brackets denote symmetry and anti-symmetry, 
respectively, under exchange of flavors between them. Each term in the 
right-hand side (r.h.s.) of Eq.~(\ref{eq:4-quark}) is again classified 
into two groups, because there are two ways to get a color-singlet 
tetra-quark $\{qq\bar q\bar q\}$ state, i.e., $\{qq\}$ 
(and $\{\bar q\bar q\}$) are of (i) $\bf{\bar 3_c}$ (and $\bf{3_c}$), 
and (ii) $\bf{6_c}$ (and $\bf{\bar 6_c}$), respectively, of color 
$SU_c(3)$. However, in the light sector ($q=u,\,d,\,s)$, these states 
would largely mix with each other through gluon exchange, because QCD 
is non-perturbative at the energy scale of these meson masses. Putting 
aside the structure of color wavefunctions, the first term 
$[qq][\bar q\bar q]$ in the right-hand-side (r.h.s.) of 
Eq.~(\ref{eq:4-quark}) seems to be well realized by the observed scalar 
nonoet, $\sigma(600)$, $a_0(980)$, $f_0(980)$~\cite{PDG06} and 
$\kappa(800)$~\cite{E791}, as predicted in Ref.~\cite{Jaffe}. 

The above light $[qq][\bar q\bar q]$ mesons have been extended to 
open-charm $[cq][\bar q\bar q]$ mesons, and, as the result, the 
charm-strange $D_{s0}^+(2317)$ has been successfully assigned to the 
iso-triplet $\hat F_I^+\sim [cn][\bar s\bar n]_{I=1}$ with 
$\bf{\bar 3_c}\times\bf{3_c}$ as the lowest charm-strange scalar 
tetra-quark meson~\cite{Terasaki-D_s}. (Our notation of the open-charm 
scalar tetra-quark mesons is provided in Ref.~\cite{Terasaki-D_s}.) 
With regard to the color configuration, the 
$\bf{\bar 3_c}\times\bf{3_c}$ state would be lower than the 
$\bf{6_c}\times\bf{\bar 6_c}$, because the forces between two quarks 
(and between two antiquarks) are attractive when they are of 
$\bf{\bar 3_c}$ (and of $\bf{3_c}$), while the forces are repulsive 
when they are of $\bf{6_c}$ (and of $\bf{\bar 6_c}$)~\cite{Hori}. 
However, mixings of the $\bf{\bar 3_c}\times\bf{3_c}$ and 
$\bf{6_c}\times\bf{\bar 6_c}$ states in open- and hidden-charm mesons 
are expected to be much smaller than that of the light sector, because 
QCD would be rather perturbative at the scale of open- and hidden-charm 
meson masses. With this assignment, we can naturally understand the 
observed narrow width and decay property of $D_{s0}^+(2317)$. In 
addition, it has been discussed that the other assignments are likely to 
be ruled out~\cite{HT-isospin,ECT-talk,Lisbon}. In our assignment of 
$D_{s0}^+(2317)$, we expect existence of its neutral and doubly charged 
partners. However, they have not yet been observed, although they have 
been searched for in inclusive $e^+e^-$ annihilation 
experiment~\cite{Babar-search}. Nevertheless, this problem seems to be 
not serious, because their production in $e^+e^-$ annihilation would be  
suppressed~\cite{Lisbon} in contrast with $D_{s0}^+(2317)$, and because 
it is expected that they could be observed in $B$ 
decays~\cite{production}. 

$X(3872)$ has been observed in the $\pi^+\pi^-J/\psi$~\cite{Belle-X-rho} 
and $\pi^+\pi^-\pi^0J/\psi$~\cite{Belle-X-omega} channels with opposite 
$G$-parities. With regard to its spin-parity, $J^P=1^+$ is favored by 
its angular analysis~\cite{Belle-X-J^P}. This implies that $X(3872)$ 
consists of two (approximately) degenerate axial-vector meson states 
with different $G$-parities, as long as $G$-parity is conserved in these 
decays as in the well-known strong interactions. It has been observed 
that this can be realized by molecular states~\cite{Oset-axial} based on 
a unitarized chiral model and also by hidden-charm tetra-quark mesons 
corresponding to the last two terms of the r.h.s. of 
Eq.~(\ref{eq:4-quark})~\cite{Terasaki-X}. 

Extension of the above open-charm scalar four-quark mesons to 
hidden-charm ones is straightforward. Their flavor wavefunctions are 
listed in Table~I. Their mass values are estimated very crudely by using 
a naive quark counting with 
$\Delta_{cs}=m_c - m_s\simeq m_{\eta_c} - m_{D_s}\simeq 1$ GeV and  
$\Delta_{sn}=m_s - m_n\simeq m_{D_s} - m_{D}\simeq 100$ MeV, 
and by taking $m_{\hat F_I} = m_{D_{s0}(2317)}= 2317$ MeV as the input 
data. Supposing that the above estimate of hidden-charm meson masses is 
not very far from the true ones, isospin conserving two-body decays 
which satisfy the OZI rule~\cite{OZI} would be limited. One of these 
decays is the $\hat\delta^c \rightarrow \eta_c\pi$, as seen in Table~I. 
However, it is not obvious if $\hat\sigma^{sc}$ and $\hat\kappa^c$ can 
decay into the $\eta_c\eta$ and $\eta_c K$ final states, respectively, 
because the roughly estimated masses of $\hat\sigma^{sc}$ and 
$\hat\kappa^{c}$ are just below the corresponding thresholds. Even if 
these decays were kinematically allowed, their rates would be much 
smaller than the rate for $\hat\delta^c \rightarrow \eta_c\pi$, because 
of their small phase space volume. 
\begin{center}
\begin{table}[t]
\caption{
Hidden-charm scalar mesons,  their flavor wavefunctions, their crudely 
estimated mass values, their assumed two-body decay channels which 
conserve isospin and satisfy the OZI-rule, and their threshold energies.  
}
\vspace{3mm}
\begin{center}
\begin{tabular}
{c|c|c|c|c}
\hline
Strangeness ($S$) & $1$ & \multicolumn{3}{c}{$0$} 
\\
\hline 
$I=1$ & 
&$\hat\delta^c\sim [nc][\bar n\bar c]_{I=1}$ &
\\
\hline
$I=\frac{1}{2}$ & $\hat\kappa^c\sim [nc][\bar s\bar c]$ &  & &
\\
\hline
$I=0$ & & & $\hat\sigma^c\sim [nc][\bar c\bar n]_{I=0}$ 
& $\hat\sigma^{cs}\sim [sc][\bar c\bar s]$ 
\\
\hline  
Mass (GeV) 
& $\sim$3.4 & $\sim$3.3 & $\sim$3.3 & $\sim$3.5
\\
\hline
OZI-allowed Decay 
& $\eta_cK$ & $\eta_c\pi$ & $\eta_c\eta$ & $\eta_c\eta$

\\
\hline 
Threshold (GeV) 
& 3.48 & 3.12 & 3.53 & 3.53
\\
\hline 
\end{tabular}
\end{center}
\end{table}
\end{center}

The rate for the $A({\bf p})\rightarrow B({\bf p'})\pi({\bf q})$ decay 
is written in the form, 
\begin{equation}
\Gamma(A\rightarrow B\pi)
=\Biggl({1\over 2J_A+1}\Biggr)\Biggl({q_c\over 8\pi m_A^2}\Biggr)
\sum_{\rm spins}|M(A\rightarrow B\pi)|^2,
                                        \label{eq:rate-general}
\end{equation}
where $J_A$, $q_c$ and $M(A\rightarrow B\pi)$ are the spin of $A$, the 
momentum of the final particles in the rest frame of $A$, and the 
amplitude for the decay, respectively. To calculate the amplitude, we 
use the PCAC (partially conserved axial vector current) hypothesis and a 
hard pion approximation in the infinite momentum frame, i.e., 
$|{\bf p}|\rightarrow \infty$~\cite{hard-pion,suppl}. (This is an 
innovation of the old soft-pion technique.) In this approximation, the 
amplitude is evaluated as 
\begin{equation}
M(A\rightarrow B\pi) \simeq \Biggl({m_A^2 - m_B^2\over f_\pi}
\Biggr)\langle{B|A_\pi|A}\rangle,   
                                                \label{eq:hard-pion}
\end{equation}
at a slightly unphysical point, i.e., $m_\pi^2\rightarrow 0$, where  
$A_\pi$ is the axial counterpart of isospin. The {\it asymptotic  
matrix element} of $A_\pi$ (matrix element of $A_\pi$ taken between   
single hadron states with infinite momentum),   
$\langle{B|A_\pi|A}\rangle$, gives the dimensionless coupling strength.   

Now we study the $\hat\delta^c \rightarrow \eta_c\pi$ decay by 
comparing with the $\hat F_I^+ \rightarrow D_s^+\pi^0$. To this aim, 
we relate asymptotic matrix elements of $A_\pi$ between hidden-charm 
meson states to those between open-charm meson states, for example, 
$\langle{D_s^+|A_{\pi^0}|\hat F_I^+}\rangle$ by using the asymptotic 
flavor $SU_f(4)$ symmetry (the $SU_f(4)$ symmetry of asymptotic matrix 
elements), 
\begin{eqnarray}
&& 
\langle{\eta_c|A_{\pi^-}|\hat\delta^{c+}}\rangle 
=\sqrt{2}\langle{\eta_c|A_{\pi^0}|\hat\delta^{c0}}\rangle 
=\langle{\eta_c|A_{\pi^+}|\hat\delta^{c-}}\rangle      
=-\sqrt{2}\langle{\eta_c|A_{\eta^s}|\hat\sigma^{sc}}\rangle 
= \sqrt{2}\langle{D_s^+|A_{\pi^0}|\hat F_I^+}\rangle.    
                                                   \label{eq:SU_f(4)}
\end{eqnarray}
In Eq.~(\ref{eq:SU_f(4)}), we have assumed that energy scale dependence 
of wavefunction overlap is mild around and above the scale of open-charm 
meson mass, although it was drastic between the mass scales of the light 
tetra-quark mesons and the open-charm ones~\cite{HT-isospin,ECT-talk}. 
It is because the quark-gluon coupling would be (rather) perturbative in 
the region of the energy scale of open-charm tetra-quark meson mass and 
beyond, while non-perturbative at the scale of light meson masses. The 
last matrix element in Eq.~(\ref{eq:SU_f(4)}) has been estimated to be 
$|\langle{D_s^+|A_{\pi^0}|\hat F_I^+}\rangle|_{SU_f(4)}\sim 0.09 - 0.13$ 
from 
$\Gamma(\hat F_I^+\rightarrow D_s^+\pi^0)_{SU_f(4)}\simeq 5 - 10$ MeV 
in Refs.~\cite{HT-isospin,ECT-talk}. Adopting the above value of the 
asymptotic matrix element as the input data, we obtain 
\begin{equation}
\Gamma(\hat\delta^{c} \rightarrow \eta_c\pi)_{SU_f(4)} 
\sim 5 - 10 \,\,{\rm MeV},                \label{eq:rate-delta^c-SU(4)}
\end{equation}
where the overlap of spatial wavefunctions is still in the $SU_f(4)$ 
symmetry limit. However, it is not yet known how to estimate 
deviation of the overlap from its $SU_f(4)$ symmetry limit in the case 
of hidden-charm mesons, although such a deviation in the case of 
open-charm mesons has been estimated~\cite{HT-isospin,ECT-talk,TM}. 

The other two-body decays of hidden-charm mesons which are kinematically 
allowed would be OZI-rule forbidden, as long as our estimate of their 
masses is not very far from the true ones. Therefore, to search for 
hidden-charm scalar tetra-quark mesons, their radiative decays would be 
important. We study them under the vector meson dominance (VMD) 
hypothesis in the same way as the open-charm scalar mesons which have 
been studied in Refs.~\cite{HT-isospin,ECT-talk}. The amplitude for the 
radiative $S\rightarrow V\gamma$ decay is written in the form  
\begin{equation}
M(S\rightarrow V\gamma)
=F_{\mu \nu }(\gamma)G^{\mu \nu }(V) A(S\rightarrow V\gamma),  
                                          \label{eq:amp-SVgamma}
\end{equation}
where $S$, $F_{\mu\nu}$ and $G_{\mu\nu}$ denote the parent scalar meson, 
field strengths of photon ($\gamma$) and vector meson ($V$), 
respectively. The amplitude for $S\rightarrow V\gamma$ decay is given by 
\begin{equation}
A(S\rightarrow V\gamma)
=\sum_{V'=\rho,\,\omega,\,\phi,\,\psi}
{X_{V'}(0)\over m_{V'}^2}A(S\rightarrow VV')
                                                   \label{eq:SVV'}
\end{equation}
under the VMD, where $X_V(0)$ is the $\gamma V$coupling strength on the 
photon-mass-shell and $A(S\rightarrow VV')$ is the $SVV'$ coupling 
strength. The values of $X_V(0)$ have been provided as 
$X_\rho = 0.033\pm 0.003$, $X_\omega = 0.011\pm 0.001$, 
$X_\phi = -0.018\pm 0.004$, $X_\psi = 0.051\pm 0.012$, 
in Ref.~\cite{Terasaki-VMD}. The OZI-rule selects a possible vector 
meson which couples to a photon. 

Related $SU_f(4)$ relation among $SVV'$ coupling strengths is given by 
\begin{eqnarray}
&&A(\hat \kappa^{c+} \rightarrow K^{*+}\psi)
=A(\hat \kappa^{c0} \rightarrow K^{*0}\psi)
=A(\hat \delta^{c+} \rightarrow \rho^{+}\psi)
=A(\hat \delta^{c0} \rightarrow \rho^{0}\psi)  \nonumber\\
&& \hspace{27mm}
=-A(\hat \sigma^{c+} \rightarrow \omega\psi)       
=-A(\hat \sigma^{sc+} \rightarrow \phi\psi)
=A(\hat \delta^{s0}\rightarrow\rho^0\phi)\beta_1,
                                           \label{eq:SVV'-SU_f(4)}
\end{eqnarray}
where $\beta_1$ is a parameter which provides an overlap of color and 
spin wavefunctions at the scale of hidden-charm meson mass. Here we take 
$|\beta_1|\simeq 1/2$ at the scale of charmed tetra-quark meson 
mass~\cite{HT-isospin,ECT-talk}, because the quark-gluon coupling is 
expected to be (rather) perturbative (around and) above the scale of 
mass of open-charm mesons. At this stage, the spatial wavefunction 
overlap is assumed to be in the $SU_f(4)$ symmetry limit, as in the
above hadronic decays. The last coupling strength in 
Eq.~(\ref{eq:SVV'-SU_f(4)}) is estimated to 
be~\cite{HT-isospin,ECT-talk}  
\begin{equation}
|A(\hat \delta^{s0}\rightarrow\rho^0\phi)|\simeq 0.02\,\,
{\rm (MeV)}^{-1}                                   \label{eq:input}
\end{equation}
from the measured rate~\cite{PDG06}, 
$\Gamma(\phi\rightarrow a_0(980)\gamma)_{\rm exp}=0.32\pm 0.03$ keV.  
\begin{center}
\begin{table}[t]
\caption{ 
Rates for radiative decays of hidden-charm scalar mesons.
} 
\vspace{2mm}
\begin{center}
\begin{tabular}
{l | c | c}
\hline
\quad Decay \quad   &  
\,\,Pole \quad   & 
\quad Rate (in MeV) \quad 
\\
\hline
$\hat \kappa^c\rightarrow K^*\gamma$ & $\psi$ & $\sim 0.23$
\\
\hline
$\hat\delta^c\rightarrow \rho\gamma$ & $\psi$  & $\sim 0.22$ 
\\ 
\hline 
$\hat\delta^c\rightarrow \psi\gamma$
& $\rho^0$ &  \hspace{2mm}$\sim 0.069$
\\
\hline
$\hat\sigma^c\rightarrow \omega\gamma$ & $\psi$  & $\sim 0.22$
\\
\hline  
$\hat\sigma^c\rightarrow \psi\gamma$ & $\omega$ 
& \hspace{2mm}$\sim 0.007$
\\
\hline
$\hat\sigma^{sc}\rightarrow \phi\gamma$ & $\psi$ & $\sim 0.24$ 
\\
\hline  
$\hat\sigma^{sc}\rightarrow \psi\gamma$ & $\phi$ &   
\hspace{2mm}$\sim 0.040$
\\
\hline
\end{tabular}
\end{center}
\end{table}
\end{center}

Inserting Eq.~(\ref{eq:amp-SVgamma}) with Eq.~(\ref{eq:SVV'-SU_f(4)})  
into Eq.~(\ref{eq:rate-general}), using the values of $X_V(0)$ listed 
above and adopting Eq.~(\ref{eq:input}) as the input data, we can 
estimate rates for radiative decays of hidden-charm scalar mesons. The 
results are listed in Table~II. As seen in the table, the rates for 
decays including $\psi$ in the final state are small because of their 
small phase space volumes. In particular, the 
$\hat\sigma^c\rightarrow \psi\gamma$ decay is more strongly suppressed 
due to the small $\gamma\omega$ coupling. Therefore, $\hat\sigma^c$ 
decays dominantly into the $\gamma\omega$ final state, and the 
$\hat\sigma^c\rightarrow \psi\gamma$ decay might not disturb the 
$\chi_{c0}\rightarrow \psi\gamma$ decay, even if the true mass of 
$\hat\delta^c$ were close to that of $\chi_{c0}$. Our results are quite 
different from those of a typical unitarized chiral 
model~\cite{Oset-rad} with the same theoretical basis as the 
one~\cite{Oset-axial} mentioned before. It is because, in this type of 
chiral model, the non-loop contribution, for example, 
$X(3700)\rightarrow ``\psi\,"\omega \rightarrow \gamma\omega$, 
where $``\psi\,"$ means a virtual $\psi$, has not been considered, 
in contrast with the present picture in which the corresponding 
${\hat\sigma^c}\rightarrow ``\psi\,"\omega \rightarrow \gamma\omega$ is 
dominant in $\hat\sigma^c$ decays as discussed above.  

So far we have assumed intuitively that $D_{s0}^+(2317)$ is the $I_3=0$ 
component of $[cn][\bar s\bar n]_{I=1}$ with 
$\bf{\bar 3_c}\times\bf{3_c}$, because the forces between two quarks 
(and two antiquarks) are attractive when they are of $\bf{\bar 3_c}$ 
(and $\bf{3_c}$). Under this assumption, we have understood its narrow 
width and its decay property~\cite{HT-isospin,ECT-talk}. Therefore, 
the low lying hidden-charm meson are expected to be of 
$\bf{\bar 3_c}\times\bf{3_c}$. To confirm that our intuitive assumption 
is feasible, two photon collision experiment would be useful. To see 
this, we denote the $I=1$ hidden-charm scalar meson with 
$\bf{6_c}\times\bf{\bar{6}_c}$ as $\hat\delta^{c*}$, and decompose 
$\hat\delta^{c}$ and $\hat\delta^{c*}$ as 
\begin{eqnarray}  
|\hat\delta^c\rangle = 
|[cn]^{\bf 1_s}_{\bf \bar 3_c} 
[\bar c\bar n]^{\bf 1_s}_{\bf 3_c}\rangle_{\bf 1_c}^{\bf 1_s}  
&&\hspace{-3mm}= 
-\sqrt{\frac{1}{4}}
{\times} {\sqrt{\frac{1}{3}}}
{|\{c\bar c\}^{\bf 1_s}_{\bf 1_c}
\{n\bar n\}^{\bf 1_s}_{\bf 1_c}\rangle_{\bf 1_c}^{\bf 1_s}}
\hspace{1mm}{+} 
{\sqrt{\frac{3}{4}}} {\times}
{\sqrt{\frac{1}{3}}}
{|\{c\bar c\}^{\bf 3_s}_{\bf 1_c}
\{n\bar n\}^{\bf 3_s}_{\bf 1_c}\rangle_{\bf 1_c}^{\bf 1_s}  }
\nonumber\\
&& \hspace{1mm} { -\sqrt{\frac{1}{4}}}
{\times}{\sqrt{\frac{2}{3}}}{|\{c\bar c\}^{\bf 1_s}_{\bf 8_c}
\{n\bar n\}^{\bf 1_s}_{\bf 8_c}\rangle_{\bf 1_c}^{\bf 1_s}}
\hspace{1mm}
{+} {\sqrt{\frac{3}{4}}} {\times}{\sqrt{\frac{2}{3}}}
{|\{c\bar c\}^{\bf 3_s}_{\bf 8_c}
\{n\bar n\}^{\bf 3_s}_{\bf 8_c}\rangle_{\bf 1_c}^{\bf 1_s} }   
                                             \label{eq:decomp-3barx3}
\end{eqnarray}
and 
\begin{eqnarray}  
|\hat\delta^{c*}\rangle = 
|[cn]^{\bf 3_s}_{\bf 6_c} 
[\bar c\bar n]^{\bf 3_s}_{\bf \bar{6}_c}\rangle_{\bf 1_c}^{\bf 1_s}  
&&\hspace{-3mm}= 
-\sqrt{\frac{3}{4}}
{\times} {\sqrt{\frac{2}{3}}}
{|\{c\bar c\}^{\bf 1_s}_{\bf 1_c}
\{n\bar n\}^{\bf 1_s}_{\bf 1_c}\rangle_{\bf 1_c}^{\bf 1_s}}
\hspace{1mm}{+} 
{\sqrt{\frac{1}{4}}} {\times}
{\sqrt{\frac{2}{3}}}
{|\{c\bar c\}^{\bf 3_s}_{\bf 1_c}
\{n\bar n\}^{\bf 3_s}_{\bf 1_c}\rangle_{\bf 1_c}^{\bf 1_s}  }
\nonumber\\
&& \hspace{1mm}{-} {\sqrt{\frac{3}{4}}}
{\times}{\sqrt{\frac{1}{3}}}{|\{c\bar c\}^{\bf 1_s}_{\bf 8_c}
\{n\bar n\}^{\bf 1_s}_{\bf 8_c}\rangle_{\bf 1_c}^{\bf 1_s}}
\hspace{1mm}
{-} {\sqrt{\frac{1}{4}}} {\times}{\sqrt{\frac{1}{3}}}
{|\{c\bar c\}^{\bf 3_s}_{\bf 8_c}
\{n\bar n\}^{\bf 3_s}_{\bf 8_c}\rangle_{\bf 1_c}^{\bf 1_s} },   
                                             \label{eq:decomp-6x6bar}
\end{eqnarray}
respectively. From the above equations, it is seen that the overlap of 
color and spin wavefunctions between $\hat\delta^{c*}$ and $\eta_c\pi$ 
states is larger by a factor $\sqrt{6}$ than that between 
$\hat\delta^{c}$ and $\eta_c\pi$ states. Therefore, if the spatial 
wavefunction overlap is not very much different from each other in both 
cases, the rate for the $\hat\delta^{c*}\rightarrow\eta_c\pi$ decay 
would be larger by about a factor 6 or more than that for the 
$\hat\delta^{c}\rightarrow\eta_c\pi$, because it is assumed that 
$m_{\hat\delta^{c*}}$ is larger than $m_{\hat\delta^{c}}$. Therefore, it 
is expected that the cross section for the 
$\gamma\gamma\rightarrow\eta_c\pi^0$ reaction would have a narrower 
($\lesssim 10$ MeV) peak at $m_{\hat\delta^c}\sim 3.3$ GeV and another 
broader peak arising from ${\hat\delta^{c*}}$ at a higher energy, if 
both of $\hat\delta^{c}$ and $\hat\delta^{c*}$ exist and 
$m_{\hat\delta^{c*}} > m_{\hat\delta^{c}}$. 
If $m_{\hat\delta^{c*}}<m_{\hat\delta^{c}}$ in contrast with our 
intuitive expectation, we need more precise information of 
$m_{\hat\delta^{c*}}$, because the gap between $m_{\hat\delta^{c}}$ and 
the $\eta_c\pi$ threshold is not very large. If 
$m_{\eta_c}+ m_\pi < m_{\hat\delta^{c*}} < m_{\hat\delta^{c}}$, the
width of $\hat\delta^{c*}$ would be sensitive to $m_{\hat\delta^{c*}}$ 
and not necessarily broader than that of $\hat\delta^{c}$, because the 
phase space volume under consideration is sensitive to it. If the mass 
of $\hat\delta^{c*}$ is lower than $\eta_c\pi$ threshold, the cross 
section would have only a peak from $\hat\delta^{c}$ around and below 
$m_{\hat\delta^{c}}$. 

In summary we have studied hidden-charm scalar $[cq][\bar c\bar q]$ 
mesons by extending the open-charm scalar $[cq][\bar q\bar q]$ mesons 
which have been studied previously and then investigated their isospin 
conserving two-body decay and radiative decays. The 
$\hat\delta^c \rightarrow \eta_c\pi$ decay will be only one two-body 
decay which conserves isospin and satisfies the OZI rule, if their 
estimated mass values are not very far from the true ones. Radiative 
decays of hidden-charm mesons also have been studied, because radiative 
channels would play an important role in search for the other 
hidden-charm scalar tetra-quark mesons. 

To confirm our intuitive expectation that the mass of 
$[cn][\bar c\bar n]$ with $\bf{\bar 3_c}\times\bf{3_c}$ is lower and 
its width is narrower than the corresponding ones with 
$\bf{6_c}\times\bf{\bar{6}_c}$, it is awaited that experiments measure 
the cross section for the $\gamma\gamma\rightarrow\eta_c\pi$ reaction 
and find two peaks; one is around $\sim 3.3$ GeV and narrower, while 
the other is beyond it and broader. 



\end{document}